\documentclass[preprint,showpacs,preprintnumbers,amsmath,amssymb]{revtex4}

\usepackage{graphicx}
\usepackage{dcolumn}
\usepackage{bm}
\usepackage{ulem}
\DeclareGraphicsExtensions{.eps}

\begin{document}
\title{Experimental study of the spatial distribution of quantum correlations in a confocal Optical
Parametric Oscillator}
\author{M. Martinelli, N. Treps, S. Ducci, S. Gigan, A. Ma\^\i tre, C. Fabre}
\affiliation{Laboratoire Kastler Brossel, Universit\'e Pierre et
Marie Curie, case 74, 75252 Paris cedex 05, France.}

\date{\today}

\begin{abstract}
We study experimentally the spatial distribution of quantum noise
in the twin beams produced by a type II Optical Parametric
Oscillator operating in a confocal cavity above threshold. The
measured intensity correlations are at the same time below the
standard quantum limit and not uniformly distributed inside the
beams. We show that this feature is an unambiguous evidence for
the multimode and nonclassical character of the quantum state
generated by the device.
\end{abstract}
\maketitle
\section{Introduction}
To date, almost all the experiments investigating the quantum
properties of the states of light produced by Optical Parametric
Oscillators (OPO) have been performed through the measurement of
the \textbf{total intensity} of the generated fields, obtained by
integrating on the detector the whole wavefront intensity
distribution. Such measurements have put in evidence the squeezed
vacuum character of the output of a degenerate OPO below threshold
\cite{Kimble}, the quantum intensity correlations between the
signal and idler beams (twin beams)\cite{Mertz}, and the bright
squeezing of the output of an optical parametric amplifier (OPA)
\cite{Schiller} and of the pump beam reflected by an OPO
\cite{Gao}.\

In the last years, the interest has turned to the spatial aspects
of quantum fluctuations, in particular because they open new
perspectives in the quantum information field : spatial features
offer the possibility of parallel processing and multichannel
operation, each part in a transverse section of a beam playing the
role of a channel. The concepts of temporal squeezing and
correlations of light beams as a whole have been extended to
spatio-temporal properties for the local quantum fluctuations in
the plane perpendicular to the propagation : squeezing of the
temporal fluctuations inside a small part of the transverse plane,
or temporal quantum correlations between different transverse
areas of the beam are some of the manifestations of the transverse
quantum properties of light. It has been shown that, in order to
obtain strong spatial quantum effects, one needs to use
\textbf{multimode nonclassical states of light}, in contrast to
the single mode operation of the experiments quoted above. It has
also been shown that such nonclassical multimode states can be
used to improve the optical resolution \cite{Optical Resolution}
and to measure small transverse displacements of a light beam
below the
standard quantum limit \cite{Small Disp,Prl}. \\
Our investigations of the spatial distribution of multimode states
of light produced by OPOs are related to the pioneer studies of
the stochastic spatial noise performed in gas lasers
\cite{laserStochast}, and more recently in semiconductor lasers
(diode lasers and VCSELs) \cite{diodeMode}, for which the spatial
distribution of the quantum fluctuations of the output beam has
been studied and interpreted in terms of a superposition of
several Hermite-Gauss modes. They must also be connected to the
studies of spatial quantum effects for the photons generated by
spontaneous parametric down-conversion, at the photon counting
level, for example to the recent demonstration of spatial
antibunching \cite{Monken}. \\
Parametric interaction in a non-linear crystal has been
theoretically shown to be a very efficient source of multimode
nonclassical states of light. Many theoretical studies
\cite{Kolobov, KolobovSokolov} have shown in particular that
Optical Parametric Amplifiers (OPA) generate multimode squeezed
states. OPOs have also been shown to be a source of multimode
squeezed states provided that they use optical cavities with
degenerate transverse modes : in particular, detailed theoretical
investigations have been performed on sub-threshold OPOs in planar
\cite{LugiatoGatti,GattiLugiato}, or quasi-planar
cavities \cite{LugiatoMarzoli,Spherical OPO}. \\
From an experimental point of view, such cavities, which are close
to instability, are quite difficult to handle and have very high
oscillation thresholds. Confocal cavities are much easier to
operate, and still exhibit interesting nonclassical transverse
effects, such as multimode squeezing in the degenerate case below
threshold \cite{Confocal Sqz}. The multimode transverse operation
of confocal OPOs above threshold has been theoretically
investigated at the classical level \cite{Transv OPO}, and
experimentally put in evidence in \cite{Matthias}, where it has
been shown that the field emitted by a confocal OPO above
threshold can be described as a superposition of a great number of
$TEM_{pq}$ modes. The quantum description of this regime is
unfortunately a difficult task, as the device switches from an
uncoupled regime of the different transverse modes to a strongly
coupled regime when
passing from the sub-threshold emission to the regime of intense output.\\
This paper is devoted to the study of the spatial distribution of
quantum fluctuations of such a device, and shows that it
generates multimode nonclassical states of light. In section II,
we give the precise definition of multimode quantum states of
light. We present in section III a criterion unambiguously
characterizing such states in our experimental configuration,
namely the measurement of the spatial distribution of the
intensity correlations between the signal and idler beams. Section
IV presents the experimental setup and the procedure for measuring
these correlations. Finally, section V presents the experimental
results proving that the signal and idler beams emitted by the
confocal OPO are multimode and spatially quantum correlated.

\section {Intrinsic characterization of multimode and single mode beams}
\subsection{Definition of single mode and multimode beams}
In the studies on optical patterns, a beam is said to be multimode
(in the $TEM_{pq}$ basis), when the far field patterns appear to
be different from the near field patterns. Actually, this feature
is a proof that the beam consists of a superposition of several
$TEM_{pq}$ modes. The electric field positive frequency envelope
of such a light beam, $E^{(+)}(\vec{r},z)$, expanded in the basis
of $TEM_{pq}$ modes $u_{p,q}(\vec r,z)$, writes :
\begin{equation}
E^{(+)}(\vec{r},z)=\sum_{p,q}\alpha_{p,q}u_{p,q}(\vec r,z).
\end{equation}
with more than one $\alpha_{p,q}$ coefficient different from zero.
Nevertheless, if the $\alpha_{p,q}$ coefficients are fixed (i.e.
if we deal with a \textbf{coherent superposition of modes} and not
a statistical one), one can always define a new mode $v_0(\vec
r,z)$ :
\begin{equation}
\label{v0}
v_{0}(\vec r,z)=\frac{E^{(+)}(\vec{r},z)}{\int\mid{E^{(+)}(\vec{r},z)}\mid^{2}d^{2}r}
\end{equation}
and construct a new orthonormal basis of modes $\{v_{i}(\vec
r,z)\}$ in which $v_{0}$ is the first element. In this new basis,
the field which appeared multimode in the Gauss-Laguerre basis, is
proportional to $v_{0}$ and is then single mode. This simple
reasoning at the classical level seems to show that the single or
multimode character of a beam having a well defined and fixed
amplitude distribution depends on the choice of the basis, and is
not an intrinsic property. A $TEM_{00}$ laser beam is single mode
in the Laguerre-Gauss basis, and multimode in the basis of
transverse plane waves. We will show here that this statement is
no longer true when one describes the beam at the quantum level.\\
We will now define a single mode quantum state of light in the
following way : $\mid\Psi>$ is a single mode quantum state of
light if there exists a basis of modes $\{v_{i}(\vec r,z)\}$ in
which it can be written as
\begin{equation}
\label{single}
\mid\Psi>=\mid\Psi_{0}>\otimes\mid{0}>\otimes\mid{0}>\otimes...
\end{equation}
where the first transverse mode $v_{0}$, whatever its shape, is a
non-vacuum state $\mid\Psi_{0}>$, and all the other modes are in
the vacuum state. We will call \textbf{intrinsic multimode states}
all states defined by a ket $\mid\Psi>$ which cannot be written as
(\ref{single}) in any basis. We will give in the next subsection a
characterization of such single mode or multimode states which is
independent of the basis used to describe it.\\
This striking difference between the classical and quantum
description of a multimode state comes from the fact that a
quantum state gives information on the mean electric field, but
also on the statistical distribution of quantum fluctuations : a
single mode beam is characterized by a well defined transverse
variation which carries all its transverse information.
Consequently, in such a state the transverse distribution of
quantum fluctuations can be deduced from the transverse variation
of the mode itself. In contrast, in intrinsic multimode states,
the spatial distribution of fluctuations and correlations cannot
be deduced from the structure of the mean field.\\
Let us finally mention that this problem is modified if one
considers \textbf{stochastic fields}, i.e. having classical
fluctuations. In this case also, the comparison between the
spatial (or temporal) distribution of the mean field and of the
classical fluctuations is still an intrinsic tool to determine
whether the field can be considered as single mode or multimode
\cite{diodeMode}. We will not consider further this problem here.

\subsection{Characterization of a single mode quantum state}
Let us consider a single mode state of light $\mid\Psi>$, written
in the adapted basis $\{v_{i}\}$ as (\ref{single}), and let us
call $\hat{a}_{i}$ the annihilation operator of photons in the
mode $v_{i}$. Let us now introduce any other mode basis of the
transverse plane $\{w_{j}\}$, and the corresponding annihilation
operators $\hat{b}_{j}$. There is a unitary transformation
relating the two basis, and correspondingly the two sets of
annihilation operators :
\begin{equation}
\label{transfo}
\hat{b}_{i}=\sum_{j}U_{ij} \hat{a}_{j}
\end{equation}
From equations (\ref{single}) and (\ref{transfo}), one deduces
that :
\begin{equation}
\label{single2}
\hat{b}_{i}\mid\Psi>=U_{i0}\hat{a}_{0}\mid\Psi_{0}>
\end{equation}
We thus obtain a specific property of a single mode quantum state
: the action on it of all the annihilation operators of any given
basis of transverse modes gives vectors which are all
proportional. One can show that this feature is a necessary and
sufficient condition to be a single mode state \cite{TrepsThese}. In
contrast, for an intrinsic multimode state, there exists at least one basis of
modes in which this property is not true. The condition for single mode states being
 very restrictive, the set of single mode states is a very small subset of the general Hilbert space.
  The complementary of this subspace, namely the set of intrinsic multimode states, is therefore much larger.\\
This characterization of single and multimode states appears to be
quite mathematical, and seems difficult to implement
experimentally. We give in the following a more convenient
property of single mode states, which is not a necessary and
sufficient condition, but which can be submitted to an
experimental check with our set-up.
\section {Characterization of single mode and multimode twin beams}
\subsection{Partial measurement of intensity fluctuations on a single mode beam}
Let us consider an intensity, or photon number, measurement using
a detector of variable transverse area $S_{A}$. We have shown in
reference \cite{Small Disp} that if the light is in a single mode
state, the variance of the photon number fluctuations measured
with this partial detector is given by :
\begin{equation}
\label{partialdetection} \Delta N_{A}^2 =
<N_{A}>+\frac{<N_{A}>^{2}}{<N_{tot}>^{2}}(\Delta
N_{tot}^2-<N_{tot}>)
\end{equation}
where $<N_{A}>$ is the mean number of photons detected on the area
$S_{A}$, $<N_{tot}>$ the mean number of photons detected on the
whole transverse plane, and $\Delta N_{tot}$ the corresponding
variance. One sees that the intensity noise normalized to shot
noise, $\frac{\Delta N_{A}^2 }{<N_{A}>}$ varies linearly with the
quantity $T=\frac{<N_{A}>}{<N_{tot}>}$. This formula can be
understood by considering that in a single mode state the photons
are randomly distributed in the transverse plane, so that a
partial detection will introduce sorting noise in the detection of
the photon number, exactly like when one introduces a linear loss
of value $T$ in front of the detector. This intuitive picture is
no longer true for multimode beams.

\subsection{Partial measurement of intensity correlations on single mode twin beams }
In a non degenerate OPO, the emitted signal and idler fields
present strong intensity correlations at the quantum level (``twin
beams"). In particular, the variance of the difference $N_{dif,tot
}=N_{1,tot}-N_{2,tot}$ between the signal (labeled 1) and idler
(labeled 2) intensities detected over the whole transverse plane
is smaller than $<N_{1,tot}>+<N_{2,tot}>$, which is the shot noise
for the sum of the signal and idler beams. When the beams produced
by the OPO are both single mode, one can easily show from
\cite{Small Disp} that the following formula, similar to
(\ref{partialdetection}), holds in case of a partial
photodetection of signal and idler beams by two detectors having
the same areas and positions $S_{A}$ in the two beams :
\begin{equation}
\label{partialdetectiondif} \Delta N_{dif,A}^2 =
(<N_{1,A}>+<N_{2,A}>)+ 
\frac{(<N_{1,A}>+<N_{2,A}>)^2}{(<N_{1,tot}>+<N_{2,tot}>)^2}(\Delta
N_{dif,tot}^2-(<N_{1,tot}>+<N_{2,tot}>))
\end{equation}
Similarly to the interpretation of formula
(\ref{partialdetection}), this formula can be understood by
considering that in single mode twin beams there is no spatial
correlations between the twin photons inside the signal and idler
beams, the photons being randomly distributed inside the two
beams. If, instead of small detectors, we use detectors with a
broad enough area to detect the whole beams, but preceded by a
diaphragm, or iris, of variable transmission $T$, expression
(\ref{partialdetectiondif}) shows that in single mode twin beams
the relative noise on the intensity difference, $n_A=\frac{\Delta
N_{dif,A}^2}{<N_{1,A}>+<N_{2,A}>}$, is a linear function of the
transmitted mean intensity. In contrast, a nonlinear variation of
this quantity will be a signature for intrinsic multimode twin
beams, for which the twin photons are not randomly distributed
inside the two beams.
\subsection{Partial measurement of intensity correlations on single mode twin beams undergoing different losses }
\begin{figure} [htbp]
\includegraphics[clip=,width=10cm]{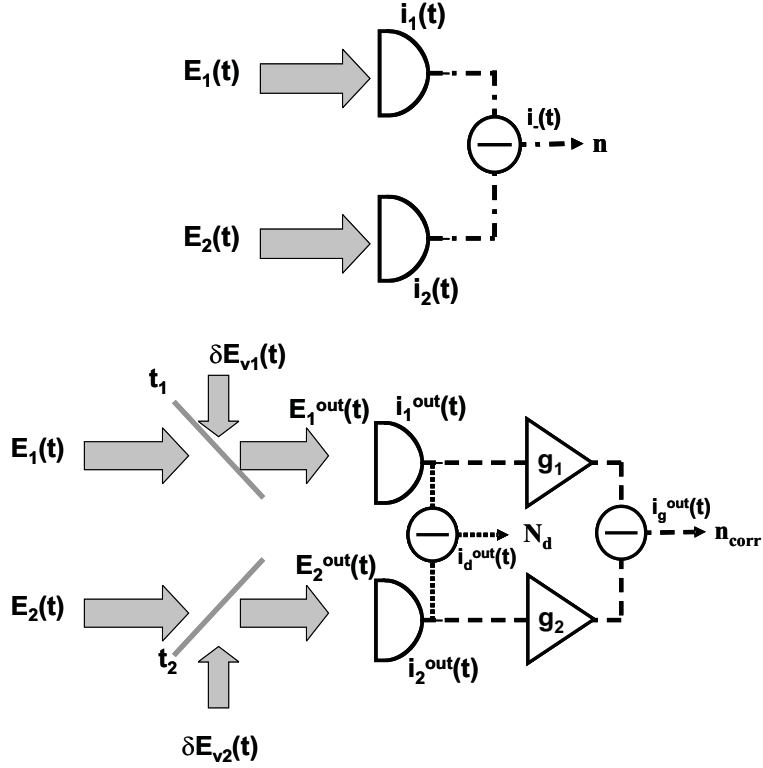}
\caption{measurement of twin beams correlation in the imbalanced
case. $i_d^{out}$: direct difference measurement. $i_g^{out}$:
difference measurement after gain corrections. } \label{unbal}
\end{figure}

We have assumed so far that the detected mean intensities of the
output beams are always equal, even in partial photodetection. In
real experiments, this is not exactly the case, and imbalances of
several percents are commonly measured, especially in confocal
OPOs, where the patterns observed in the signal and idler beams
are different. The imbalance will affect the intensity
fluctuations of the two beams in a different way. The imbalance of
the whole beams is due to the fact that the signal and idler
beams, produced in exact equal amounts in the nonlinear crystal,
experience different losses in their propagation. Furthermore, in
the case of a partial photodetection, one must also take into
account that the signal and idler beams may have different shapes.
It is important to precise the way to define the ``twin character"
of such imbalanced beams in a partial measurement. We will first
introduce the normalized intensity difference noise $n$ when the
whole beams are measured (Fig.\ref{unbal}a). We will then consider
a partial measurement (Fig.\ref{unbal}b) and define the direct
normalized intensity difference noise $n_d$, with no corrections,
as it can be directly calculated from the measurement. We will
finally define the corrected normalized intensity difference noise
$n_{corr}$, an attempt
to recover the value of $n$ from $n_d$ assuming that the beams are single mode.\\
Assuming that the mean signal and idler fields are real, we can
write the field intensity operator, in the small fluctuation
approximation, as :
\begin{equation}
\delta \hat{i}_{m}=2E_{m}\delta \hat{E}_{m}
\quad m\in
\{1,2\}
\end{equation}
where $\delta \hat{E}_{m}$ is the quadrature fluctuation operator
in the amplitude direction. We now introduce the normalized
quantity $n$, which is the variance of the intensity difference
fluctuations normalized to the total shot noise, and can be
written as :
\begin{equation}
n=\frac{\langle(\delta \hat{i}_{1}-\delta \hat{i}_{2})^{2}\rangle}
{4(i_{1}+i_{2})\Delta^{2}E_{v}}
\label{twin}
\end{equation}
where $\Delta^{2}E_{v}$ is the noise variance of the vacuum field.
$n>1$ for beams with classical correlations (obtained for example
by using beamsplitters) and $n<1$ for quantum correlated beams
(``twin beams").
Let us now assume that the two beams are subject to losses before
being incident on the photodetectors (Fig.\ref{unbal}b). We call
$E_m^{out}$ the fields which are incident on the photodetectors
and $i_m^{out}$ their intensities, and $n_d$ the normalized
variance of their intensity difference :
\begin{equation}
n_{d}=\frac{\Delta ^2 i_d^{out}}{4(i_{1}^{out}+i_{2}^{out})\Delta^{2}E_{v}}=\frac{\langle(\delta
\hat{i}_{1}^{out}-\delta
\hat{i}_{2}^{out})^{2}\rangle}
{4(i_{1}^{out}+i_{2}^{out})\Delta^{2}E_{v}}
\end{equation}
In the transverse single mode case, losses of any origin and
partial photodetection have the same effect on the fluctuations
mechanism, identical to the effect of a beam splitter of
transmission $t_{m}$ (m=1,2) (with $t_m^2=i_m^{out}/i_m $). The
field intensity $\hat{i}_{m}^{out}$ measured by the photodetectors
present fluctuations $\delta \hat{i}^{out}_{m}$(t) given by:
\begin{equation}
\delta \hat{i}^{out}_{m}(t)=2t_{n}^2 E_{m} \delta \hat{E}_{m} + 2
t_{m} \sqrt{(1-t_{m}^2)}E_{m} \delta \hat{E}_{vm} \quad m\in
\{1,2\} \label{outIntensity}
\end{equation}
Knowing the effective loss coefficient $t_{m}$, it is possible to
adjust the gain $g_{m}$ on the acquisition channels at the value
$1/t_{m}^{2}$, as shown in fig. \ref{unbal}, so that the mean
values of the corrected intensities $g_{m}\langle
\hat{i}^{out}_{m}\rangle$ are equal to their value without losses
$\langle \hat{i}_{m}\rangle$. If one assumes that the two beams
are single transverse mode, it is possible to recover the original
normalized intensity difference noise $n$ of eq. (\ref{twin}) by
subtracting the effect of the vacuum fluctuations $\delta
\hat{E}_{vm}$. Writing $\Delta^{2} i_g^{out}$ the fluctuations of
the difference between the two acquisition channels :
\begin{equation}
\Delta ^2 i_g^{out} = \langle (g_{1}\delta
\hat{i}^{out}_{1}-g_{2}\delta \hat{i}^{out}_{2})^2\rangle
\label{variance}
\end{equation}
One easily shows that the quantity
\begin{equation}
n_{corr}=\frac{\Delta ^2 i_g^{out}}{4
(g_{1}i^{out}_{1}+g_{2}i^{out}_{2})\Delta ^2 E_{v}}
+1-\frac{g_{1}^2 i^{out}_{1}+g_{2}^2
i^{out}_{2}}{g_{1}i^{out}_{1}+g_{2}i^{out}_{2}} \label{sqzCorr}
\end{equation}
is equal to $n$ in the single transverse mode case. The two last
terms of (\ref{sqzCorr}) correspond to the corrections arising
from vacuum fluctuations introduced by the losses or the partial
detection.

\subsection{Multimode twin beams}
Since for single mode signal and idler beams $n_{corr}=n$, a constant value $n_{corr}$ with the transmittance
$t_{m}$ for a partial measurement (or a linear variation of $n_d$)
is a good indication (but not a proof) of the single mode
character of the signal and idler mode beams. In contrast a non
constant value of $n_{corr}$, or a nonlinear variation of $n_d$,
with respect to the losses is a an unambiguous signature of the
multimode character of the two beams generated by the OPO.
Furthermore, if these quantities are smaller than 1, we can
conclude that we are in presence of multimode nonclassical beams.
Strictly speaking, as the vacuum correction in $n_{corr}$ has been
derived assuming single mode beams, $n_{corr}$ do not anymore
correspond exactly to a noise correlation in the case of multimode
beams. Moreover, for a partial measurement, the mean intensity of
both beams may be different, and neither $n_d$ nor $n_{corr}$ are
perfectly suited for the exact characterization of quantum
intensity correlation. For these reasons, in the case of a
\textbf{partial} measurement of a \textbf{multimode} beam, more
sophisticated criteria should be investigated, that will be
developed in a forthcoming publication.

\section {Experimental setup}
For OPOs operating in cavities with degenerate transverse modes,
the formation of complex spatial structures has been theoretically
\cite{Lugiato, Maxi} and experimentally \cite{Matthias,Sara Ducci}
studied. Furthermore, quantum multimode operation of OPO has been
already theoretically predicted \cite{LugiatoGatti}. But, up to
now, to our knowledge, there has been no experimental
demonstration that the emission of an OPO is intrinsically
multimode. The experiment presented here investigates the spatial
distribution of the intensity correlation between the signal and
idler beams emitted by a confocal OPO above threshold in order to
identify their single mode or multimode character.
\begin{figure} [htbp]
\includegraphics[clip=,width=15cm]{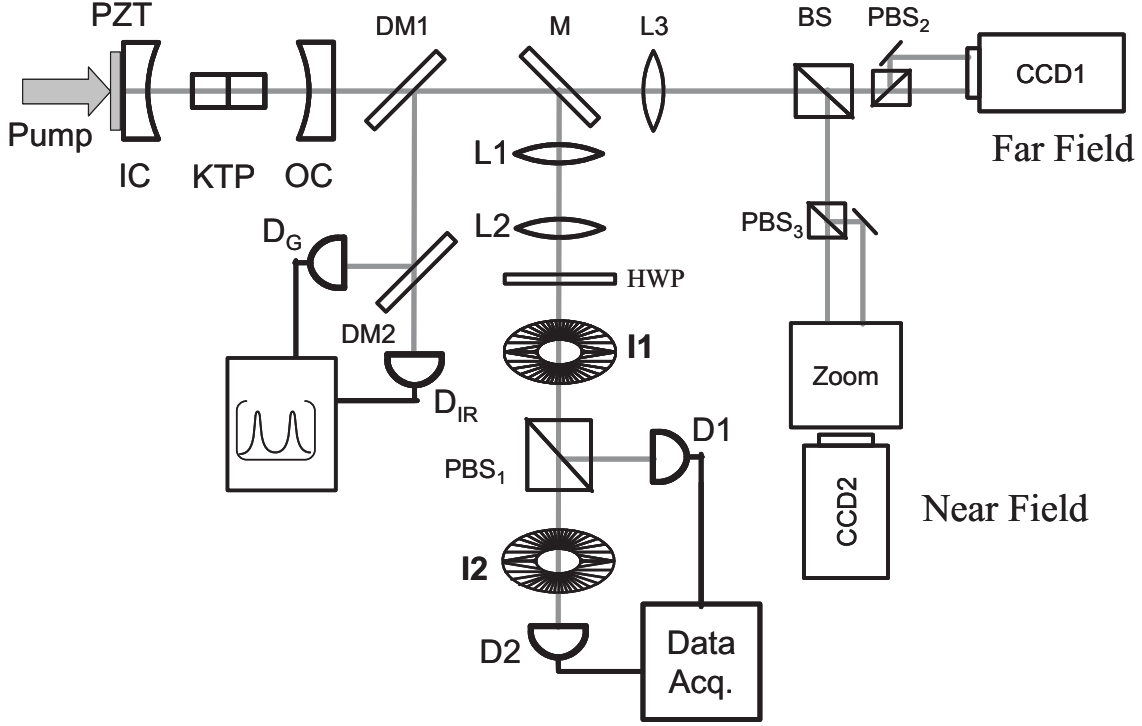}
\caption{Experimental setup}
\label{manip}
\end{figure}
The setup, shown in Fig.\ref{manip}, can be divided in two parts:
the triply resonant OPO and the acquisition system of the
intensity fluctuations of the beams.
\subsection{The triply resonant OPO}
In order to match the desired requirements of power and stability
of the experiment, a dedicated setup was made for the generation
of the pump beam for the OPO. We used an ultrastable single mode
Nd:YAG (yttrium aluminum garnet) laser with 350 mW of output power
to lock a flash-lamp-pumped Nd:YAG laser operating in a ring
cavity. This configuration \cite{Matthias,Sara Ducci} gives a
stable single mode beam with an output power of 3.5 W at 1064 nm.
This beam is injected in a semi-monolithic frequency doubler, using a
MgO:LiNbO$_{3}$ crystal and a concave mirror. The output of the
second harmonic generator reproduces the frequency stability of
the injected Nd:YAG laser and gives a total output power of 1.3 W
at 532 nm. The pump is then injected inside the OPO with a waist
equal to twice the $TEM_{00}$ waist of the OPO cavity mode like in
reference \cite{Matthias}.\\
The nonlinear crystal used inside the OPO is a walk-off
compensated KTP (potassium titanyl phosphate) cut for type II
phase matching. It is made by two 5 mm long crystals that are
optically contacted, with their orientations compensating the
walk-off effect. The triply resonant OPO uses two plane-concave
mirrors, whose curvature is R = 100 mm. The
input coupler (IC) has a transmittance of 10\% for the pump (532
nm) and high reflectance for the Nd:YAG wavelength (1064 nm). The
output coupler (OC) has a high reflectance at 532 nm and 1\%
transmittance at 1064 nm. Measured finesse value of the cavity
with the crystal is 40 at 532 nm and 300 at 1064 nm and temperature is
controlled such that signal (o-polarization) and idler (e-polarization)
emitted by the OPO are close to frequency degeneracy. The cavity length can be
tuned over a few free spectral ranges with a piezo-electric
ceramic (PZT) attached to the input coupler. A coarse control of
the length of the OPO is performed by means of translation stages
fixed on both cavity mirrors. A fraction of the infrared output of
the Nd:YAG laser can also be injected into the cavity for
alignment purposes and to check the transverse mode separation.
The OPO works close to the confocal configuration. In order to
define the region of confocality, it is useful to remember that
the distance L between two spherical mirrors for a confocal cavity
depends not only on the ray of the concave surface but also on the
refractive index of the medium inside the cavity. Therefore,
considering the diffraction effects inside a crystal of length
$\ell$ and refractive index $n$ in a cavity made by two spherical
mirrors of ray $R$, the distance L between the mirrors for the
confocality will be \cite{cavityLength}
\begin{equation}
L_{conf}=R+\ell (1-1/n) \label{Lcav}
\end{equation}
Since the refractive index for pump, signal and idler are
different (respectively 1.7881, 1.8296 and 1.7467
\cite{xtalData}), the length of the cavity $L_{conf}$ for which
the exact confocality is obtained, is different for each of the
three modes (104.41, 104.28 and 104.53 mm). Transverse degeneracy
is achieved when different transverse modes are resonant for the
same cavity length. Due to the width of the resonance peaks,
transverse degeneracy can be obtained even if the cavity is not
exactly confocal. The range of confocality can be defined as the
region of the cavity length where the fundamental and the first
transverse even mode separation is smaller than the cavity
bandwidth, and, following \cite{siegman}, can be expressed as
\begin{equation}
|L-L_{conf}|<\frac{\pi R}{2 F} \label{DeltaL}
\end{equation}
where F is the finesse of the cavity. In our case, the range of
confocality for signal, idler (and pump) is larger than the
difference between the confocal lengths for the signal, idler and
pump modes. Transverse degeneracy, essential for multimode
emission, is then achievable at the same time for the three modes.
Taking the average of the confocal length for signal and idler, we
will consider the confocal length as $L_{conf}=104.41 mm$, and
express the cavity length in terms of $\Delta L=L-L_{conf}$. \\
Although the threshold for oscillation is quite low and around
30mW, significant thermal effects take place inside the cavity
even close to threshold, because of the non-negligible absorption
of the green light(3\% at 532nm) and of infrared light (0,45\% at
1064nm) : thermal lensing changes the total Gouy phase shift added
to the wave in a round trip \cite{siegman} and even a sub-confocal
cavity can become transverse degenerate if the pump power injected
into it is sufficiently high. Therefore the confocality
\cite{Matthias} is obtained for cavity lengths which are
shorter than the confocal length
$L_{conf}$, defined above for a "cold" OPO with no thermal effects.
\subsection{Detection and acquisition}
The details of the setup used for studying the transverse
distribution of noise in twin beams cavity can be seen in
Fig.\ref{manip}. Near the output coupler, a dichroic mirror (DM1)
is used to eliminate the residual green light transmitted through
the OPO cavity. A small fraction (2\%) of the infrared light that
is reflected by this mirror and transmitted by a second dichroic
mirror (DM2) is monitored by an InGaAs infrared photodetector
$D_{IR}$. It is used to stabilize the OPO cavity by a servo-loop
made by a high-voltage amplifier connected to the PZT, controlling
the output power of the IR beam and stabilizing it during a time
ranging from seconds to minutes of continuous operation. The green
light reflected by DM1 and DM2
is detected by a visible photodetector $D_G$.\\
After the dichroic mirror (DM1) at the OPO output, a flipping
mirror (M) allows either the imaging of signal and idler far and
near fields on a screen, or the recording of their intensities by
two InGaAs infrared four-quadrants photodetectors D1 and D2
(ETX505Q from Epitaxx). In this experiment, only one quadrant of
each photodetector is used. Their quantum efficiency are very
close (less than 1\% difference) and equal to 90\% $\pm$ {5\%}. In
both configurations, polarizing beamsplitters (PBS) separate the
signal and idler beams. Depending of the orientation of the half
wave plate (HWP) placed before PBS1, signal and idler can be sent
either to D1 or D2. During the measurement process, two irises of
variable diameter are used to select in the far
field a circular region of the output beam. Iris I1 is
used to select a narrow circular region on both signal and idler,
while iris I2 acts only on a single beam, either signal or idler.
The lenses L1 and L2 are adjusted for each experiment in order to
project the far field of the beam into the iris plane. The
transmitted intensities are recorded and their fluctuations are
monitored by a dedicated data acquisition system
that is described below.\\
The data acquisition system used for the noise measurement differs
from the usual method of direct measurement and subtraction of
noise fluctuations (e.g. Ref.\cite{twinOPO}). The high frequency
part of the photocurrent of each photodetector is amplified by a
transimpedance amplifier and a broadband 36 dB amplifier. This
signal is then mixed in an electronic demodulator at a frequency
$f_0$ equal to 3.5 MHz which lies inside the cavity bandwidth for
the infrared modes and above the excess noise frequencies of the
output beams. The output of the mixer has an active low-pass
filter, working at 100 kHz. This signal is registered by a fast
Analog-to-Digital acquisition card for measuring the noise
correlation of the photocurrents. We used two acquisition cards
(PCI6110E from National Instruments) with four simultaneous
measurement channels each and 12 bits for signal conversion. The
signal is measured with a repetition rate of 200 kHz and
registered in the computer. Remaining channels of the data
acquisition system are used for the measurement of the average
value of the photocurrent of each detector, as well as the
photocurrents of detectors $D_G$ and $D_{IR}$. From the stored
information of the average value and noise fluctuations of the
photocurrent we can calculate the noise correlation of the
intensity, and compare it to a previous calibration of the shot
noise level made with a single output of the OPO or with the
injected IR light from the Nd:YAG laser. This technique allows us
to acquire in a very short time interval both the temporal
fluctuations of a given Fourier component of the photocurrents and
of their mean values, and then to postprocess the stored data. We
can thus determine the different normalized quantities that we
have defined earlier in that paper.

\subsection {Experimental procedure}
When the OPO oscillation is stabilized for a given cavity length,
we perform two kinds of experiments. In the first experiment, we
record the simultaneous noise distribution of the signal and idler
beams by using the iris I1. We continuously close the iris I1 in
2s, while acquiring a long series of values of $\delta i_s$,
$\delta i_i$, $i_s$, $i_i$, $i_{IR}$ and $i_G$ (6 series of
400,000 simultaneous values). We analyze the data by groups of
10,000 values, calculating for each group the average photocurrent
value $\langle i_m \rangle$ (m=1,2) and the normalized variances
$n_{corr}$ and $n_d$. The corresponding vacuum fluctuation
$\Delta^2 E_v$ was previously calibrated, and the electronic noise
level is subtracted in the calculation of the variances. The
transmittance of the iris for each series of 10000 points is
defined as $T= \frac{r}{r_{open}}$ where $r$ is the ratio
$r=\frac{\langle i_s+i_i \rangle}{\langle i_{IR}\rangle}$, and
$r_{open}$ corresponds to the initial value of $r$ when the
diaphragm is open. The normalized intensity difference noise
variances $n_{corr}$ and $n_d$ are then plotted as a function of
the transmission of the iris.\\
The second kind of experiments consists in setting iris $I_2$ on
either the signal or the idler beam path and taking the unaffected
beam as a reference. The procedure is identical to the previous
one, except that $r$, for the iris on the signal beam, is defined
as $r=\frac{\langle i_s\rangle}{\langle i_{IR}\rangle}$. Closing
the iris $I_{2}$ attenuates only one of the beams, and produces a
strong imbalance between the photocurrents of the two
photodetectors, with a ratio ($i_s/i_i$) well out of the range
90\% to 110\%, which was typical in the previous conditions. With
such a large imbalance, the value of $n_{d}$ deduced from the
experiment does not give any quantitative information on the
quantum correlations between the two beams. On the other hand, the
calculation of the corrected value of the normalized noise
$n_{corr}$ allows one to recover the information on the quantum
correlation only in the case where the two beams are single mode.
Therefore, in this second kind of experiments, we only calculate
$n_{corr}$ as a function of the transmittance $T$.
\section {Spatial distribution of the intensity correlation}
\subsection {Experiments with the iris on both signal and idler beams}
\begin{figure}[htbp]
\includegraphics[clip=,width=8cm]{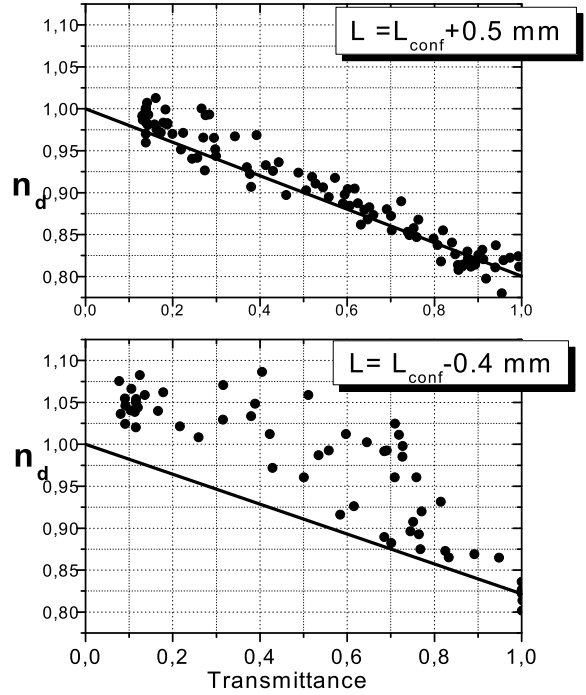}
\caption{spatial distribution of the normalized intensity
difference noise $n_d$ as a function of the transmittance T with
iris $I_1$ on signal and idler beams. Points: values of $n_d$ for
a cavity length equal to $L_{conf}=L+0,5mm$ (beyond the
confocality range)and $L_{conf}=L-0,4mm$ (inside the confocality
range). Straight line: Values of $n_d$ that would be obtained with
a single mode beam having the same squeezing when the iris is
fully open.} \label{demo}
\end{figure}

As described in the previous subsection, only iris $I_1$ is
present. We have realized in this configuration two series of
measurements: in the first one, the OPO oscillation is stabilized
for a cavity length outside the confocality range, ensuring a
$TEM_{00}$ output mode for signal and idler beams
(ref.\cite{Matthias, Sara Ducci}) ; in the second one, the cavity
length is set to a shorter value, inside the confocality range,
where complicated patterns can be observed in the signal and idler
beams. While closing continuously the iris $I_{1}$, we record the
transmitted intensity of the signal and idler modes and the
normalized intensity noise difference variance $n_d$. The
different normalized variances are then plotted
as a function of the transmittance $T$. In the figures
(\ref{demo}), the straight line represents the value of $n_d$ that
could be calculated with a single mode beam having the same
intensity correlation as the whole beam.
One observes that outside the confocality range, where the signal
and idler fields are close to $TEM_{00}$, the calculated straight
line fits well the experimental points. Moreover, for an open
diaphragm, the normalized intensity difference noise variance is
equal to 0.8, demonstrating significant quantum intensity
correlations between signal and idler.\\
In the confocality range, where the signal and idler fields have
more complicated, and different, transverse variations, one
observes that the variation of $n_d$ is no longer linear with the
transmittance, demonstrating that the emission of the OPO is
multimode. Moreover, when the iris is open in that confocal
configuration, the noise correlation is still below shot noise and
close to the value obtained for the single mode emission: the
quantum intensity correlations for the whole beams are preserved
even for multimode beams. The increase of noise when the iris is
closed shows that the intensity correlation is stronger in the
outer parts of the beams than in the center. Unfortunately, with
such an experiment it is difficult to describe more precisely the
transverse distribution of correlations between the signal and
idler beams.
\subsection {Experiments with the iris either on signal or idler beam}
\subsubsection{Single mode beams}
\begin{figure} [htbp]
\includegraphics[clip=,width=10cm]{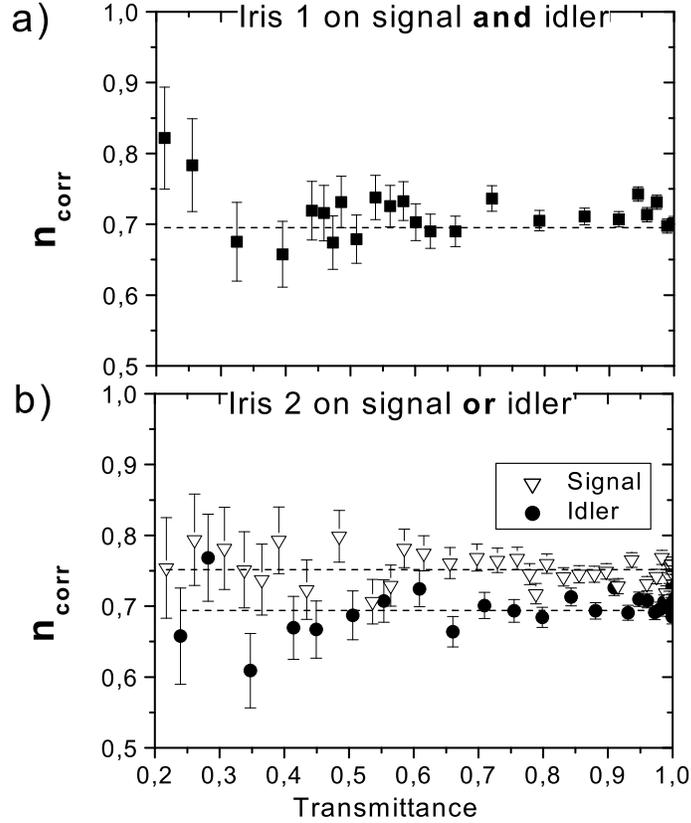}
\caption{ Experimental values of $n_{corr}$ for quasi-$TEM_{00}$
signal and idler beams. a) Iris on both beams b)Iris on a single
beam. Idler ($\bullet$) and signal($\circ$). Dashed lines: average
corrected noise.} \label{example}
\end{figure}

For a length of cavity $L=L_{conf}+5,6mm$, where the OPO is
supposed far from transverse degeneracy, we compared the results
of the two kinds of experiments for the values of $n_{corr}$ as a
function of the transmittance. The results are plotted in
Fig.\ref{example}.
For all those experiments, as the beam intensity on the
photodetector is reduced, the fluctuations of the photocurrent
approaches the electronic level. The dispersion in the calculated
data increases when we close the iris. Typically, the variance of
the intensity noise reached the dark noise level for an incident
power of 0.2 mW, obtained with a typical iris transmittance of
10\%.\\
One observes that all the experimental points are aligned on a
horizontal straight line, with a mean value around 70\%, which
shows that the signal and idler beams are single mode,
quantum-correlated, beams. The difference in the obtained level of
noise for in fig \ref{example}b (69\% and 75\%) is certainly due
to the fact that both experiments are performed at different
moments, and therefore on possibly different longitudinal modes.
The values of $n_{corr}$ remain stable during a
single series of measurements when we close the iris.
\subsubsection {Beams emitted by an OPO close to confocality}
\begin{figure} [htbp]
\includegraphics[clip=,width=18cm]{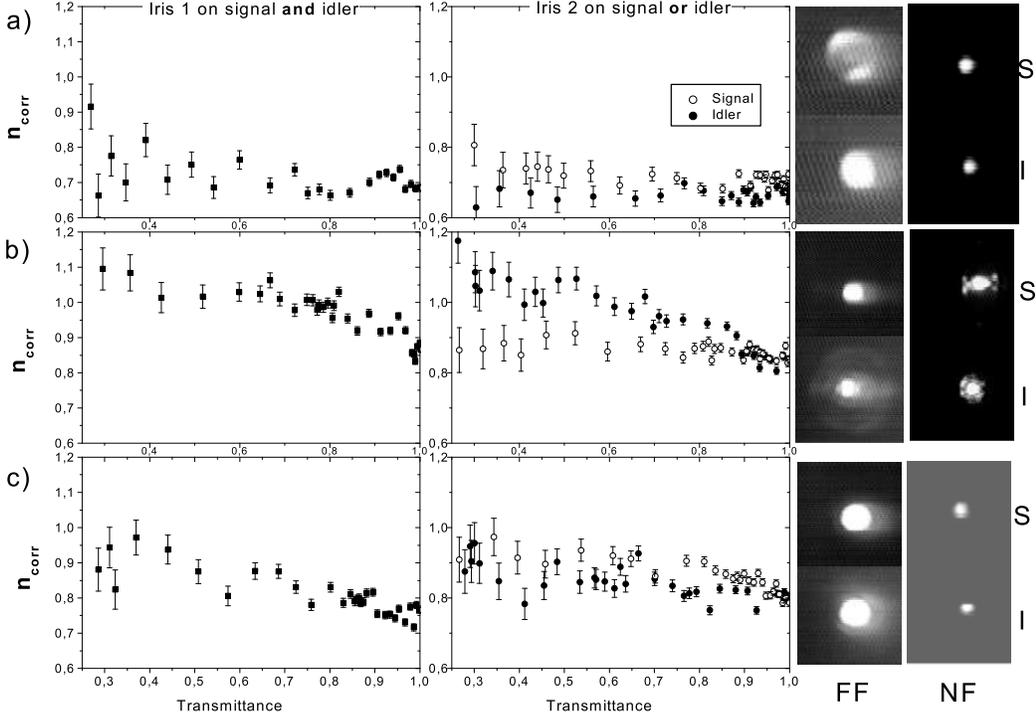}
\caption{Experimental values of $n_{corr}$ for signal and idler
beams displaying complex transverse distributions. The distance
from confocality ($\Delta L$) is equal respectively to a) 0.38, b)
-0.37 , and c) -0.62 mm. Left, iris on both beams:
($\blacksquare$). Center, iris on a single beam. Idler ($\bullet$)
and signal($\circ$). Right: Far field (FF) and near field (NF)
images of signal and idler beams.} \label{result}
\end{figure}

The length for the cavity is chosen to be very close to the exact
confocal length $L_{conf}$. In this configuration complex
structures appear in far and near fields of the signal and idler
beams \cite{Matthias}. We performed the same experiments and
analysis as in the previous subsection, adding the image of the
near field and the far field of the beam obtained with the CCD
cameras. Fig.\ref{result}a, b and c display respectively the
results obtained for $L-L_{conf} = 0.38, -0.37$ and $-0.62 mm$.
In the configuration of Fig.\ref{result}a, one observes a nearly
gaussian distribution of the idler beam and a non-gaussian
distribution of the signal beam. When the iris $I_2$ is placed on
the idler beam, $n_{corr}$ does not deviate in a clear way from a
flat line, except for lower values of the iris transmittance. In
this situation, the dispersion in the results increases, following
the attenuation of the idler field intensity. When we set the iris
on the signal path, the corrected noise deviates slightly from a
constant value for values of the transmittance close to 50\%. The
same behavior can be observed when we perform the experiment with
iris 1 on both signal and
idler. \\
In the configuration of Fig.\ref{result}b, more complex structures
appear : a ring pattern can be seen in the outer part of the idler
far field image. In the near field, complex structures appear for
both signal and idler beams. In this situation, when iris $I_1$ is
put on both beams, the corrected correlation noise $n_{corr}$ goes
up to the shot noise level and stabilizes around this value. It is
interesting to observe that the level of squeezing obtained is
reduced, and the normalized noise for open iris starts at 85\%.
When the iris $I_2$ is put on the signal path and closed,
$n_{corr}$ remains constant: from these observations, one can
infer that the photons in the signal beams which are correlated to
the idler photons are uniformly distributed inside the signal
beam. On the contrary, when the diaphragm $I_2$ is set on the
idler beam, the plot of $n_{corr}$ deviates from a flat line. In
that case the results are very close to those obtained when the
iris $I_1$ is set on both signal and idler, showing that most of
the correlated photons in the idler are concentrated in the
external part of the beam. The center part of the idler is
constituted essentially of non-correlated photons. \\
In the configuration of Fig.\ref{result}c, both signal and idler
beams present a faint external ring out of the central maximum. In
that case, the evolution of the noise when closing iris 1 or
closing iris 2 for signal or idler are very close. Both signal and
idler corrected noises $n_{corr}$ are reduced when one closes iris
I2, showing that unlike the centers of signal and idler beams,
their outer parts are quantum correlated.
We see that these experiments give interesting indications on the
details of transverse distribution of the correlations inside the
beams. Unfortunately, the great quantity of modes oscillating
simultaneously in a confocal cavity, which makes a theoretical
approach of the system very difficult, and the lack of long term
stability of the setup, which induces a large dispersion on the
experimental results, prevent us from a more quantitative
comparison between theory and experiments, which has been tackled
in the case of the multimode VCSEL for example \cite{diodeMode}.
\section{Conclusion}
The results presented in this paper show that the intensity
correlations in a confocal OPO are at the same time below the
standard quantum limit and not uniformly distributed inside the
beams, which is a clear evidence that the quantum state generated
by such an OPO is a multimode nonclassical state of light. This
experiment showed also that even in the multimode case, the
intensity fluctuations of the signal and idler beams remain
quantum correlated. It would be very interesting to find a
theoretical explanation, even qualitative, to the fact that, in
our experimental configuration, the central part of the beams seem
to be less quantum correlated than the outer parts.
We plan to extend this study in the case of an OPO working in a
lower degeneracy cavity, where only some few modes are allowed to
oscillate simultaneously. The small number of oscillating modes
can lead to a detailed theoretical description of the system.
Another interesting regime is the operation below threshold, where
the comparison with theory is somehow simpler, but which requires
a spatially resolved homodyne detection. For a more general point,
this demonstration of an intrinsic multimode emission of signal
and idler beams in the quantum regime of the above threshold OPO,
opens interesting prospects for the study and use of transverse
quantum effects in bright beams.
\begin{acknowledgments}
Laboratoire Kastler-Brossel, of the Ecole Normale Sup\'{e}rieure
and the Universit\'{e} Pierre et Marie Curie, is associated with
the Centre National de la Recherche Scientifique.\\
This work was supported by the European Commission in the frame of
the QUANTIM project (IST-2000-26019).\\
M. Martinelli, on leave from the Instituto de Fisica, Universidade
de S\~{a}o Paulo, PO Box 66318 CEP, S\~{a}o Paulo-SP Brazil,
wishes to thank Coordena\c c\~{a}o de Aperfei\c coamento de
Pessoal de N\' \i vel Superior (CAPES-BR) and Funda\c c\~{a}o de
Amparo \`{a} Pesquisa do Estado de S\~{a}o Paulo (FAPESP-BR) for
funding. Agn\`es Ma\^\i tre and Sara Ducci are also at the
p\^{o}le Mat\'eriaux et ph\'enom\`enes quantiques FR CNRS 2437
Universit\'e Paris 7 Denis Diderot.
\end{acknowledgments}

\end{document}